\documentclass[aps,prl,twocolumn,superscriptaddress]{revtex4-1}

\usepackage{amsmath,amssymb,graphicx}
\usepackage{url}
\usepackage[colorlinks,urlcolor=blue]{hyperref}

\begin{document}

\title{Viewpoint: Vector meson spin alignment by the strong force field}

\author{Xin-Nian Wang}
\email[]{Email: xnwang@lbl.gov}
\affiliation{Nuclear Science Division MS 70R0319, Lawrence Berkeley National Laboratory, Berkeley, California 94720, USA}
\affiliation{Physics Department, University of California, Berkeley, California 94720, USA}
\begin{abstract}
{\bf Observation of unexpectedly large global spin alignment of $\phi$ vector mesons in non-central heavy-ion collisions by STAR experiment may reveal the non-perturbative nature of quark interaction in hot matter through fluctuating strong force field with short correlation length.}
\end{abstract}
\maketitle

\begin{figure}
\centerline{\includegraphics[scale=0.23]{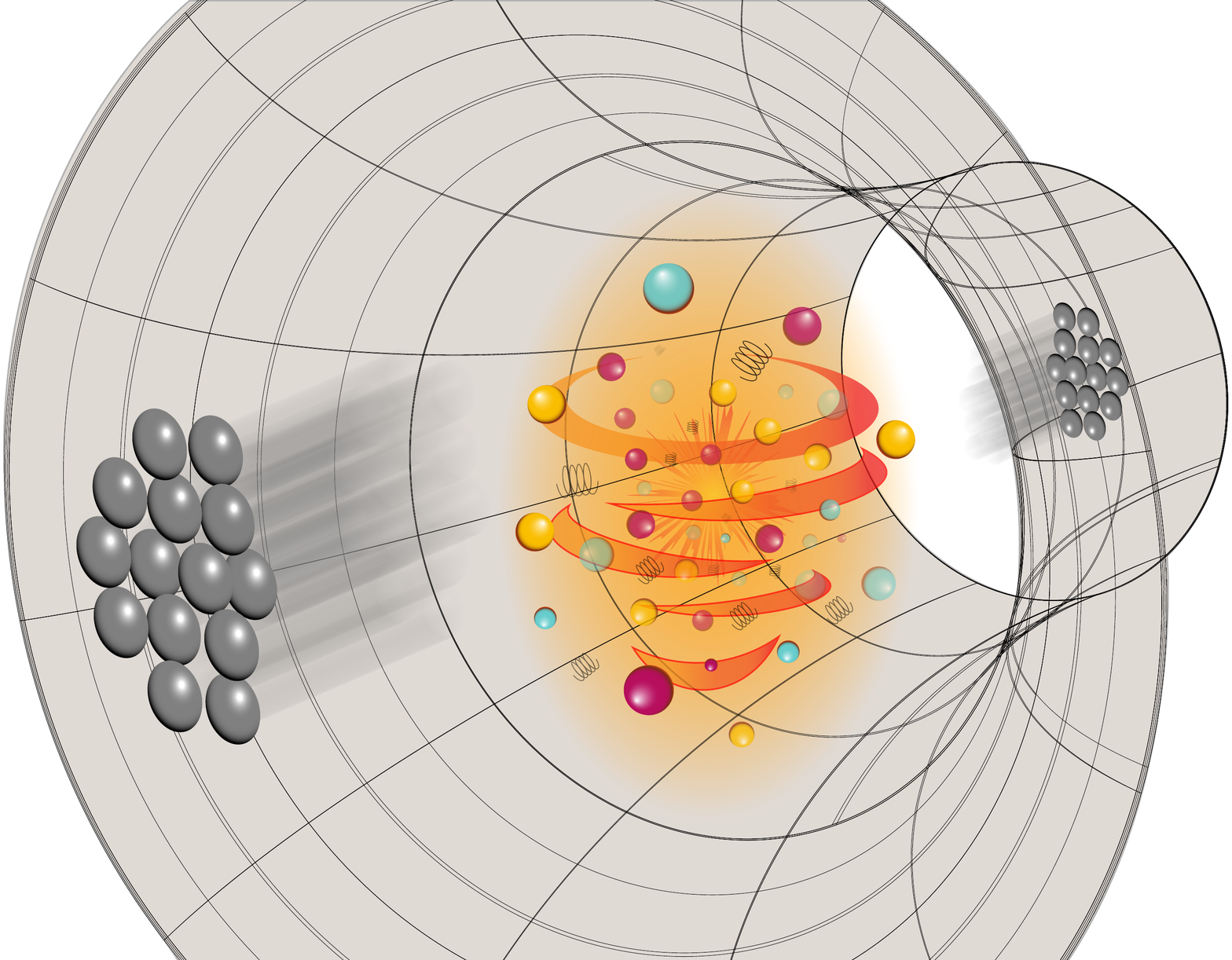}}
\caption{An illustration of the vorticity field in the overlap region of non-central heavy-ion collisions.}
\label{fig:1}
\end{figure}

\begin{figure}
\centerline{\includegraphics[scale=0.35]{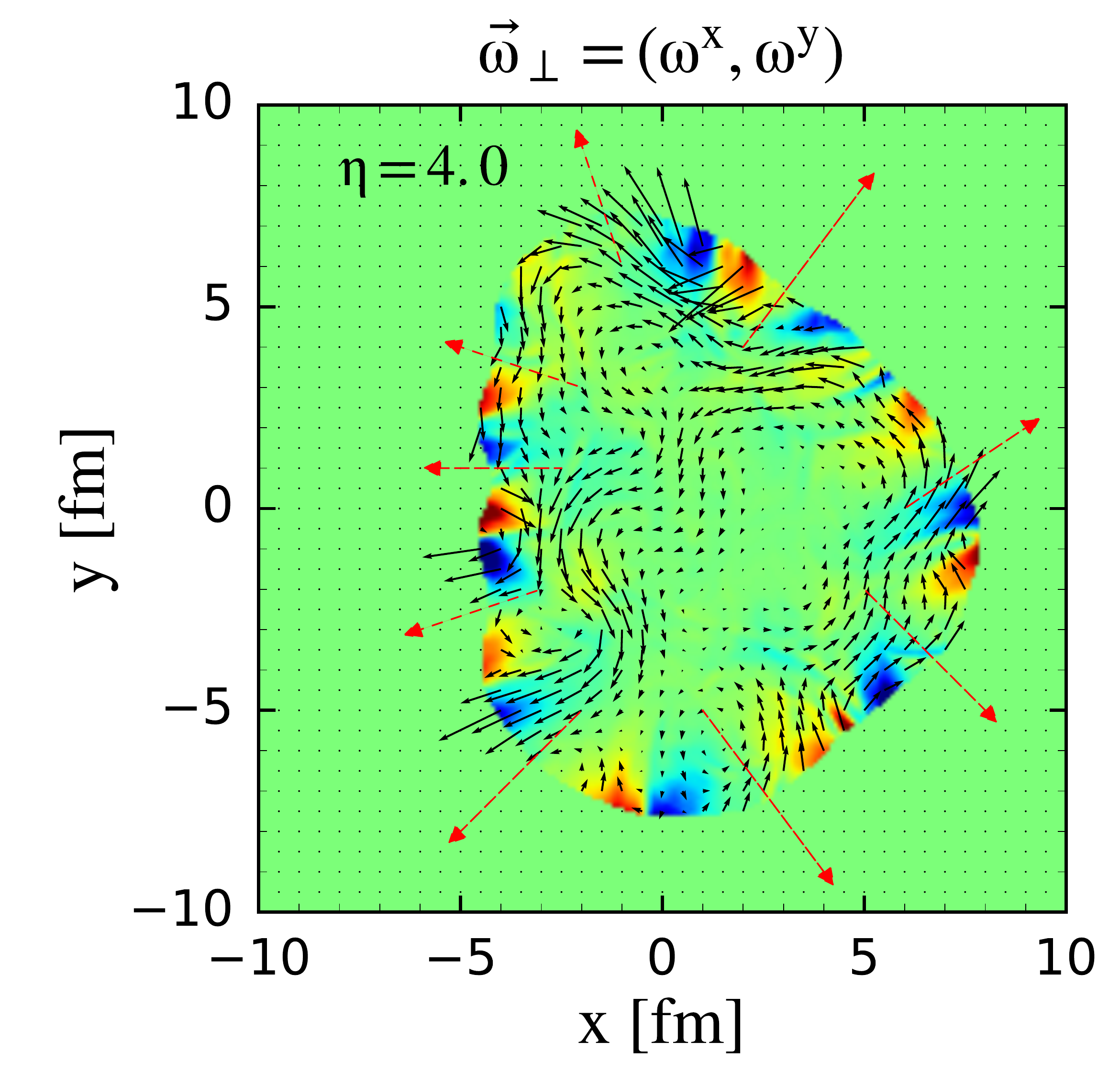}}
\caption{The transverse vorticity distribution in a noncentral Au+Au collision at RHIC \cite{Pang:2016igs}.}
\label{fig:vort}
\end{figure}

In non-central heavy-ion collisions, the system carries a large amount of orbital angular momentum in the order of $10^3\times (p_{in}/{\rm GeV}) \hbar$ that is proportional to the beam momentum $p_{in}$ per nucleon in the center of mass frame\cite{Liang:2004ph,Gao:2007bc}. At low energies, such collisions produce highly deformed compound nuclei with large spins \cite{Vo:1993zz}. In collisions at the Relativistic Heavy-ion Collider (RHIC)  and the Large Hadron Collider (LHC) energies, a new form of matter called quark-gluon plasma (QGP) is formed in which quarks and gluons can roam freely across the whole volume of the matter instead of the domain of a nucleon. Such a new state of matter is predicted by the lattice QCD calculation \cite{HotQCD:2014kol} to have an equation of state (EoS) with a rapid cross-over phase transition that is much softer as compared to that of a compound nucleus. This soft EoS is indeed supported by a Bayesian analysis of the existing data on soft hadrons \cite{Pratt:2015zsa}. The large orbital angular momentum in these collisions therefore cannot give rise to a rotating QGP. Instead, only a small fraction of the total orbital angular momentum is transferred to the dense matter in the form of transverse gradient of the longitudinal flow velocity or transverse vorticity as illustrated in Fig.~\ref{fig:1} and shown in Fig.~\ref{fig:vort} from hydrodynamic model simulations. 

Such a transverse vorticity in the GQP fluid was referred to as the local orbital angular momentum and predicted by Liang and Wang \cite{Liang:2004ph} to lead to the global spin polarization of the QGP in non-central heavy-ion collisions along the opposite direction of the reaction plane. One of the consequences of the global quark polarization is the global spin polarization of the final state hyperons such as $\Lambda$ and $\bar\Lambda$. In a constituent quark model, the spin of $\Lambda$ ($\bar\Lambda$) is carried by the strange quark (anti-quark). The quark polarization due to spin-orbital coupling will lead to the same global polarization of $\Lambda$ ($\bar\Lambda$) and the polarization of $\Lambda$ and $\bar\Lambda$ are approximately the same.   More than a decade later, this predicted phenomenon was indeed observed through the measurement of global spin polarization of the final-state $\Lambda$ and $\bar\Lambda$ hyperons in STAR experiment at RHIC beam-scan (BES) energies \cite{STAR:2017ckg}. Assuming thermal equilibrium in spin degrees of freedom for the produced hyperons and given the freeze-out temperature, the measured global polarization 1-2\% indicates a vorticity $\omega\approx 9\times 10^{21}$ per second. This is the most vortical fluid observed in nature. In the meantime, the QGP is also found to behave like a perfect and strongly coupled fluid with a small shear viscosity to entropy ratio approaching to the uncertainty bound $1/4\pi$ \cite{Bernhard:2019bmu}. It is also opaque to energetic jets of quarks and gluons leading to the suppression of large transverse momentum jets and hadrons \cite{Qin:2015srf}. The experimental data from both RHIC and LHC experiments therefore point to the formation of the hottest, most perfect, opaque and vortical fluid in nature. The local vorticity and therefore the final spin polarization increases with decreasing beam energy and become sizable at the RHIC BES energies.

In their follow-up studies, Liang and Wang also predicted vector meson spin alignment \cite{Liang:2004xn} due to the same mechanism for the hyperon global spin polarization. Since a vector meson with spin 1 can have three different spin orientations, the probability for its spin to align with a given direction, for example the reaction plane of heavy-ion collisions, is 1/3. Any value of the spin alignment probability different from 1/3 means the polarization of the vector mesons along that direction.
Unlike a hyperon whose spin is carried by that of a single strange quark in a constituent quark model and therefore its polarization is linear in vorticity, the spin of a vector meson comes from both of its constituent quark and anti-quark and its polarization (deviation of the alignment probability from 1/3) is therefore quadratic in local vorticity. This spin alignment in Au+Au collisions at the highest RHIC energy was first explored by Jinhui Chen and Yu-Gang Ma within STAR collaboration back in 2005 \cite{STAR:2008lcm} without conclusive observation because of the limited statistics. 
Encouraged by the sizable hyperon spin polarization observed at RHIC BES energies, the team recently renewed their effort on the measurement of the vector spin alignment at these lower beam energies. They indeed observed for the first time the spin alignment of $\phi$ and $K^{*0}$ vector mesons \cite{STAR:2022fan}. The analysis was mainly carried out by a joint team of Fudan University, Institute of Modern Physics of CAS, Brookhaven National Laboratory,  University of Illinois at Chicago, and Kent State University, led by Jinhui Chen, Declan Keane, Yu-Gang Ma, Subhash Singha, Xu Sun, Aihong Tang and Chensheng Zhou. Though the measured spin alignment for $K^{*0}$ is consistent with zero, it is, however, $2\sim 3$ orders of magnitude larger for $\phi$ mesons than that caused by the vorticity of the fluid as extracted from the  global hyperon polarization and electro-magnetic field in the colliding system. Spin alignments caused by other effects are also estimated to be negligible. 

To explain such unexpectedly large spin alignment of $\phi$ vector mesons in non-central heavy-ion collisions, a recent study by X. L. Sheng, L. Oliva, Z. T. Liang, Q. Wang and X. N. Wang \cite{Sheng:2022wsy} proposed a quark polarization mechanism by the strong force field. In this model, quarks interact with the dense medium through a strong force and become polarized similarly as they do under the influence of electromagnetic field. The strength of the strong force field can be much stronger than the electromagnetic field and the coupling is expected to be two orders of magnitude larger. Such mechanism can therefore lead to very large spin alignment of vector mesons. The strong force field is assumed to fluctuate and flavor-dependent with a short range correlation. It therefore will not contribute to the global spin polarization of hyperons but lead to the spin alignment of flavor singlet vector mesons which is proportional to the short distance (in the range of a hadron size) correlation of the field strength. Since there is no correlation between the strong force fields for different quark flavors, it will not lead to spin alignment of vector mesons with different quark and ant-quark flavors such as $K^{*0}$.   Within this model, STAR Collaboration extracted from their measurement of $\phi$ spin alignment the strong field fluctuation strength . It is about 2 times that of $\pi$ mass squared $m_\pi^2$. Given the strong field strength, the final $\phi$ meson spin alignment will depend on the details of the quark coalescence model of QGP hadronization, for example, the coalescence coupling constant and the hadron wave-function over which the strong force field correlation is averaged.  Once these uncertainties are known or reduced, one can potentially extract the correlation strength of the fluctuating strong force field in the QGP and shed new light on the nature of non-perturbative interaction between quarks and gluons at high temperature and density. The strong force correlation will provide a set of new information about the short distance structure of QGP and the nature of QCD phase transition.

%About the author:
\vspace{0.5cm}

{\bf Xin-Nian Wang} is currently a Senior Scientist at Lawrence Berkeley National Laboratory, an Exceptional PI Researcher at University of California at Berkeley and a Fellow of American Physical Society. His main research interest is in high-energy particle and nuclear physics, especially in the search for a new form of matter known as the Quark-Gluon Plasma in high-energy heavy-ion collisions. His recent work focuses on hard probes of QGP, relativistic hydrodynamic models and application of machine learning.

%\begin{figure}
%\centerline{\includegraphics[scale=0.15]{xnwang.jpg}}
%\end{figure}

\end{document}